\documentclass[apj]{emulateapj}
\usepackage{multirow}
\usepackage{longtable}
\usepackage{rotating}
\usepackage{color,soul}

\def\gtorder{\mathrel{\raise.3ex\hbox{$>$}\mkern-14mu
             \lower0.6ex\hbox{$\sim$}}}
\def\ltorder{\mathrel{\raise.3ex\hbox{$<$}\mkern-14mu
             \lower0.6ex\hbox{$\sim$}}}

\shorttitle{The super-fast spin of NEA (455213) 2001 OE84}

\shortauthors{Polishook et al.}

\begin{document}

\title{The fast spin of near-Earth asteroid (455213) 2001 OE84, revisited after 14 years: constraints on internal structure}

\author{D.~Polishook\altaffilmark{1},
N.~Moskovitz \altaffilmark{2},
A.~Thirouin \altaffilmark{2},
A.~Bosh \altaffilmark{3},
S.~Levine \altaffilmark{2},
C.~Zuluaga \altaffilmark{3},
S.~C.~Tegler \altaffilmark{4},
O.~Aharonson\altaffilmark{1},
}

\altaffiltext{1}{Department of Earth and Planetary Sciences, Weizmann Institute of Science, Rehovot 0076100, Israel}
\altaffiltext{2}{Lowell Observatory, 1400 West Mars Hill Road, Flagstaff, AZ 86001, USA}
\altaffiltext{3}{Massachusetts Institute of Technology, 77 Massachusetts Avenue, Cambridge, MA 02139, USA}
\altaffiltext{4}{Department of Physics and Astronomy, Northern Arizona University, Flagstaff, AZ, 86011, USA}

\begin{abstract}
At a mean diameter of $\sim650~$m, the near-Earth asteroid (455213) {\it 2001 OE84} ({\it OE84} for short) has a rapid rotation period of $0.486542\pm0.000002~$hours, which is uncommon for asteroids larger than $\sim200~$m.
We revisited {\it OE84} 14 years after it was first, and last, observed by Pravec et al. (2002) in order to measure again its spin rate and to search for changes. We have confirmed the rapid rotation and, by fitting the photometric data from 2001 and 2016 using the lightcurve inversion technique, we determined a retrograde sense of rotation, with the spin axis close to the ecliptic south pole; an oblate shape model of $a/b=1.32\pm0.04$ and $b/c=1.8\pm0.2$; and no change in spin rate between 2001 and 2016.
Using these parameters we constrained the body's internal strength, and found that current estimations of asteroid cohesion (up to $\sim80~$Pa) are insufficient to maintain an intact rubble pile at the measured spin rate of {\it OE84}. Therefore, we argue that a monolithic asteroid, that can rotate at the rate of {\it OE84} without shedding mass and without slowing down its spin rate, is the most plausible for {\it OE84}, and we give constraints on its age, since the time it was liberated from its parent body, between $2-10$ million years.


\end{abstract}


\section{Introduction}
\label{sec:Introduction}

\begin{figure}
\centerline{\includegraphics[width=17cm]{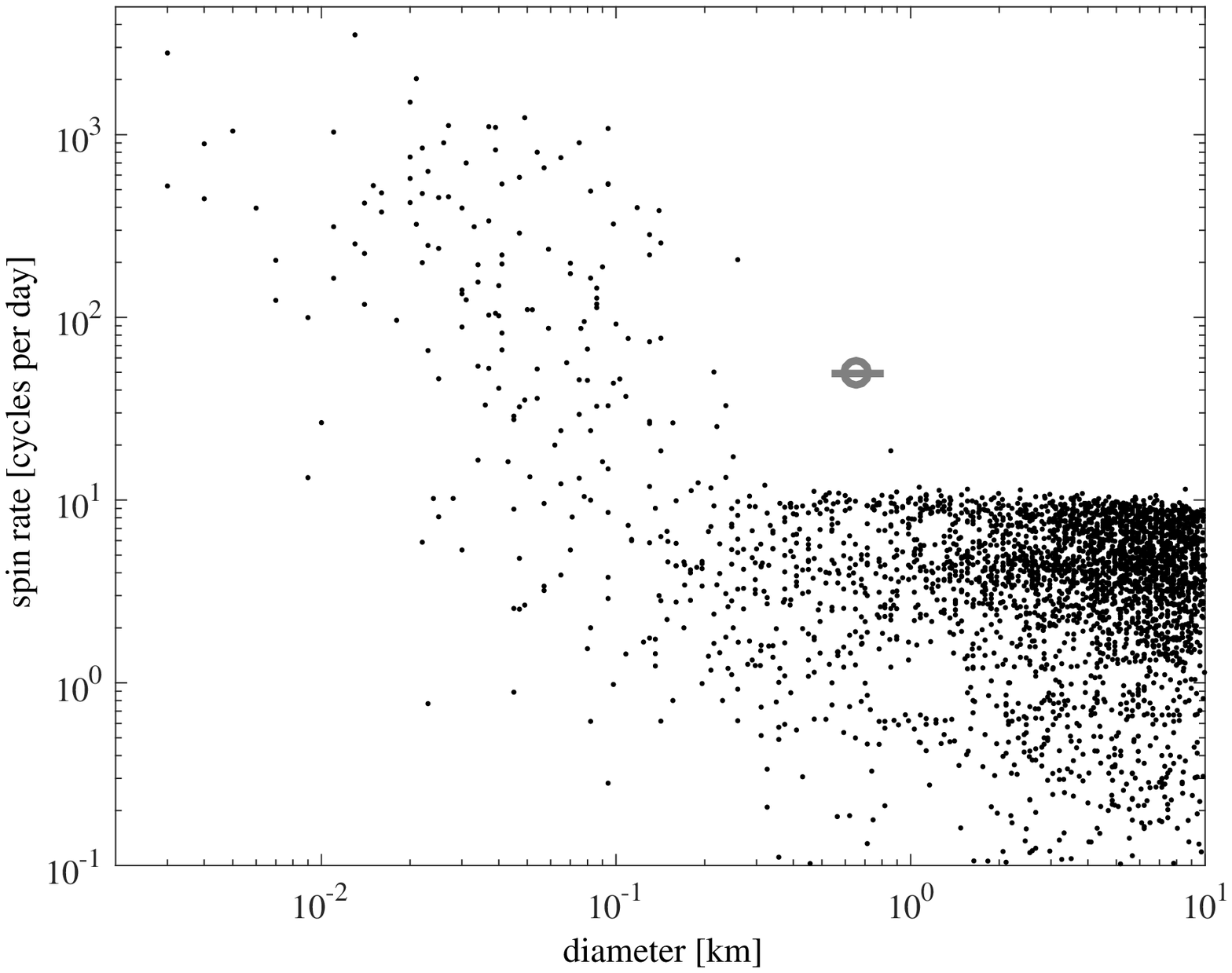}}
\caption{Asteroid diameters vs. spin rates (black dots). {\it OE84} is marked with a grey circle. The uncertainty on the diameter is marked by the grey horizontal line. The uncertainty of the spin rate is too small to see on this scale. The asteroid data is from the lightcurve database (Warner et al. 2009). The {\it spin rate barrier} at about $10$ cycles per day for asteroids larger than $\sim200~$m is easily noticed.
\label{fig:DiamSpinDiag}}
\end{figure}

The near-Earth asteroid (NEA) (455213) {\it 2001 OE84} ({\it OE84} for short) has been known for more than fifteen years as a unique object that ``defies" the spin rate barrier - the threshold at which asteroids larger than $\sim200~$m are not observed to rotate faster (Fig.~\ref{fig:DiamSpinDiag}). This spin barrier suggests that most, if not all, asteroids larger than $\sim200~$m and up to tens of kilometers, are rubble piles - conglomerates of rocks held together only by the weak force of their self gravity ({\it e.g.}, Harris 1996, Richardson et al. 1998). In this paradigm, small asteroids ($D\lessapprox200~$m) with spin rates faster than the spin barrier have strong internal structures with significant internal tensile strength.

Even though a few large asteroids (with $D>200~$m) have been found to rotate faster than the spin barrier at $\sim 2~$hours (Chang et al. 2014, 2015, 2017, Naidu et al. 2015, Polishook et al. 2016), none is as fast as the  $0.5~$hour rotation period of {\it OE84}. Pravec et al. (2002) measured its lightcurve in October through December of 2001 with sufficiently high signal to noise to unambiguously determine the anomalously high rotation rate. Furthermore, a 4-peaked lightcurve with twice the period is impossible due to a high measured amplitude of $\sim0.6~$mag  (Harris et al. 2014). Pravec et al. (2002) measured {\it OE84}'s reflectance spectrum and found it matches an S-type taxonomy. Therefore, they estimated an albedo of 0.18, which combined with an absolute magnitude $H_v=18.31\pm0.16$, suggest an effective diameter of $0.7~$km.

The unique position of {\it OE84} on the diameter-spin plane (Fig.~\ref{fig:DiamSpinDiag}), can be explained by one of the two following physical models:

1. {\it OE84} could have a rubble pile structure with sufficient internal cohesion between its components to resist centrifugal disruption. A leading theory (Scheeres et al. 2010) suggests that sub-millimeter-sized gravel has enough cohesion due to week van der Waals forces to act as a glue between meter-sized and larger boulders. Holsapple (2007) used the Drucker-Prager yield criterion, a pressure-dependent metric for determining whether a material has failed or deformed, to show that with enough cohesion, objects smaller than $\sim200~$m can rotate rapidly without being considered monolithic, while larger bodies ($D\gtrapprox10~$km) will still break up at the {\it rubble pile spin barrier}, regardless of their cohesion. Since this yield criterion is size dependent, objects in an intermediate size range ($\sim200~$m to $\sim10~$km) can rotate faster, in principal, than the currently observed spin barrier ($P\lessapprox2~$hours). However, this cohesion model cannot explain why there are so few intermediate-sized asteroids beyond the spin barrier  (Fig.~\ref{fig:DiamSpinDiag}). Since the uniqueness of {\it OE84} could arise from unusual internal cohesion, we derived the asteroid shape model to constrain the three physical axis of the geoid, $A$, $B$ and $C$ (where $A\geq B\geq C$), and thus inform the minimal cohesion needed to keep {\it OE84} bound.

2. {\it OE84} could be a large monolithic body with non-zero tensile strength (Pravec et al. 2002). In this case, {\it OE84} is at least 10-30 times larger in volume than any other presumed monolithic body. Such a scenario would be possible if {\it OE84} were a fragment of a much larger, intact body ($>100~$km) that was destroyed by a catastrophic collision. However this requires an explanation of how such a structure has remained intact while presumably other similarly sized collisional fragments have been further disrupted into smaller pieces.

Below we discuss our observations, performed 14 years and 3 months after the observations presented by Pravec et al. (2002). This is followed by analysis of the spin state and a discussion regarding the internal structure of {\it OE84}.

\section{Observations and measurements}
\label{sec:observations}

Observations were conducted over four nights on Lowell Observatory's $4.3~$m Discovery Channel Telescope (DCT) between January 19 to March 12, 2016, in excellent photometric conditions. Data from one additional night (Feb 22, 2016) was omitted from our analysis due to low signal-to-noise in the resulting photometry, primarily due to  proximity of the full moon. We employed the Large Monolithic Imager (LMI, $6k~$x$~6k$ pixels) with 3x3 binning and a field of view of $12.5'~$x$~12.5'$ (Levine et al. 2012). In order to maximize the signal to noise we used a wide-band {\it VR} filter, roughly equivalent to the combined {\it Johnson V} and {\it Cousins R} filters. Guiding the telescope at sidereal rates allowed the asteroid to stay within a single field for each observing night. Over our two months of observations, the visible magnitude of {\it OE84} ranged from 20.5 to 21.2. Observational circumstances are summarized in Table~\ref{tab:ObsCircum}.

Reduction followed standard procedures such as subtracting bias and dividing by a normalized and combined image of dome and twilight flat fields. Aperture photometry was performed with an aperture radius equal to 4 times the measured seeing (typically $\sim 1"$) and calibration of differential magnitude was achieved using hundreds of local comparison stars. Only comparison stars with $<0.02~$mag variation were used for calibration. The photometry was corrected to unity of geo- and heliocentric distances and the data were corrected by light-travel time using values for the orbital geometry from the JPL Horizons website\footnote{http://ssd.jpl.nasa.gov/horizons.cgi}. This reduction procedure is described in more detail in Polishook $\&$ Brosch (2009). In addition, we calibrated the instrumental magnitudes of field stars against the PanSTARRS catalog (Flewelling et al. 2016) assuming we were using the {\it SDSS r} filter. This calibration generally included $\sim100$ field stars per image and was achieved using an automated photometry pipeline (Mommert 2016). The net precision of this calibration (combined uncertainty in zero point and photon noise in the source photometry) was generally $<0.05$ magnitudes. This precision does not reflect uncertainty in the transform from our {\it VR} filter to {\it SDSS r}. However, this calibration did allow us to place the measured photometry on a single magnitude scale.

\section{Albedo and size estimation}
\label{sec:sizeEstimation}

The decrease in brightness as a function of the phase angle correlates with the asteroid's taxonomy and albedo (Belskaya \& Shevchenko 2000). We used the mid-peak value of the lightcurves in each night, corrected to unity of geo- and heliocentric distances, to derive the phase curve and to match a linear fit to it. The fit has a slope of $0.026\pm0.004~$mag/deg, consistent with the S-type taxonomy (Fig. 4 at Belskaya \& Shevchenko 2000). Since there is nothing suggesting {\it OE84} is any different from a typical S-type we adopt an albedo of $0.197\pm0.051$ which is statistically derived from WISE spacecraft data (Pravec et al. 2012). New data would be required to know the actual albedo value.

We could not use the phase curve to improve the accuracy of the absolute magnitude of {\it OE84} since our observations were carried out in a wide-band {\it VR} filter. Based on our field star photometry we estimate that the associated uncertainty in the transformation of {\it VR} magnitudes to a standard system (e.g. {\it V Johnson}) for sources with solar-like colors is $\sim 0.2~$mag. This translates to an error on the absolute magnitude larger than the value given by Pravec et al. (2002). Therefore, we estimate an effective diameter of $650^{+160}_{-110}~$m by adopting the absolute magnitude from Pravec et al. (2002) and the average S-type albedo mentioned above. The asteroid's additional physical parameters are summarized in Table~\ref{tab:PhysicalParam}.

\section{Spin analysis}
\label{sec:spinAnalysis}

\subsection{Synodic period}
\label{sec:synodicperiod}

We employed a Fourier series analysis with 4 harmonics to find the spin frequency that best matches the data. We searched a wide range of frequencies (8 to 88 cycles/day) at an interval of $0.00001$ cycles/day. For a given frequency, a least-squares minimization was used to derive a $\chi^2$ value. The frequency with the minimal $\chi^2$ is chosen as the most likely period. The error in the best-fitting frequency is determined by the range of periods with $\chi^2$ smaller than the minimum $\chi^2$ + $\delta\chi^2$, where $\delta\chi^2$ is calculated from the inverse $\chi^2$ cumulative distribution function at $3\sigma$ ($\sim99.7\%$) and with $1+2M$ degrees of freedom, where $M$ is the number of harmonics used in the fit. This procedure is further detailed in Polishook et al. (2012).

\begin{figure}
\centerline{\includegraphics[width=17cm]{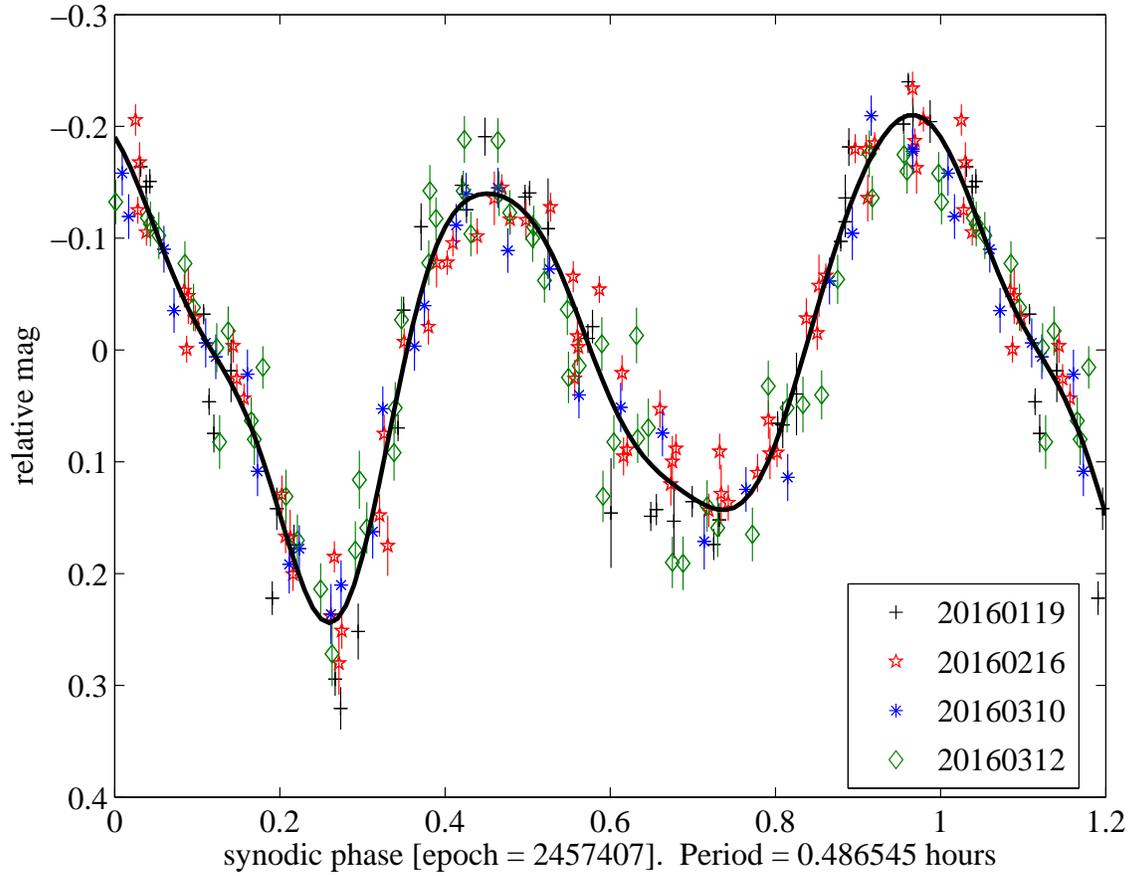}}
\caption{Folded lightcurve of {\it OE84} from our 2016 campaign. The best-fit period (solid black curve) is $0.486545\pm0.000004~$hours. See Table~\ref{tab:ObsCircum} for the observational circumstances.
\label{fig:OE84_LC}}
\end{figure}

The synodic rotation period was found to be $0.486545\pm0.000004~$hours. The lightcurve displays two unique maxima and minima with maximum amplitude of $0.45\pm0.05~$mag and minimum of $0.28\pm0.03~$mag (Fig.~\ref{fig:OE84_LC}). During our 2016 observing window, the amplitude did not change significantly since the sub-observer coordinates of the asteroid changed by merely $2^o$. In other words, the asteroid was observed from approximately the same aspect through 2016, even though the phase angle increased from $7^o$ to $19^o$. The amplitude did change by $\sim50\%$ in 2001 as the sky coordinates changed by $\sim35^o$ and the phase angle spanned a range from $17^o$ to $25^o$ (Pravec et al. 2002). Our rotational phase-folded lightcurve is presented in Fig.~\ref{fig:OE84_LC}.

To compare our derived lightcurve properties with those of Pravec et al. (2002) we run our code on their data after correcting it for light-travel time and reducing it to unity for geo- and heliocentric distances. For the 2001 data we derived a synodic period of $0.4865155\pm0.0000061~$hours which is equal to the value published by Pravec et al. (2002), though with slightly higher uncertainty (they used a uniform weighting for all of their measurements, while we did not; Harris et al. 1989). Even though the measured difference between the 2001 and the 2016 rotation periods is $\sim0.1~$seconds we will show in the following section that this is the result of the difference between the measured synodic rotation periods compared to the true sidereal period.

\subsection{Lighhtcurve inversion technique}
\label{sec:inversion}

Photometric data obtained across multiple apparitions and viewing geometries can be fit with a model that includes the sidereal period, spin axis coordinates and a convex shape. This is achieved by employing the lightcurve inversion technique developed by Kaasalainen and Torppa (2001) and Kaasalainen et al. (2001), and implemented as a software package by Josef {\v D}urech ({\v D}urech et al., 2010) and Josef Hanu{\v s} (Hanu{\v s} et al., 2011). The combination of photometric measurements from two apparitions distanced by 14 years provides sufficient data to derive a small range of statistically acceptable solutions for the sidereal period, and observing at different aspect angles constrains the shape model. See Polishook (2014) regarding the significance of results from this inversion technique and the determination of uncertainties. We used the code (provided to us by J. {\v D}urech) with and without an additional free parameter that allows for a linear change in the spin rate across the 14 years. This technique has identified three asteroids spun-up by the YORP effect (1862 Apollo, Kaasalainen et al. 2007; 1620 Geographos, {\v D}urech et al. 2008; and 3103 Eger, {\v D}urech et al. 2012).

We run the code on the entire range of the spin axis coordinates ($0^o$ to $360^o$, $-90^o$ to $90^o$) using a fixed interval of $5^o$ while matching a wide range of sidereal periods. We run the code in a few iterations, choosing the best sidereal period and then decreasing the search range and time interval for each subsequent run. An interval of $10^{-8}~$hours was used in the final iteration.

\begin{figure}
\centerline{\includegraphics[width=17cm]{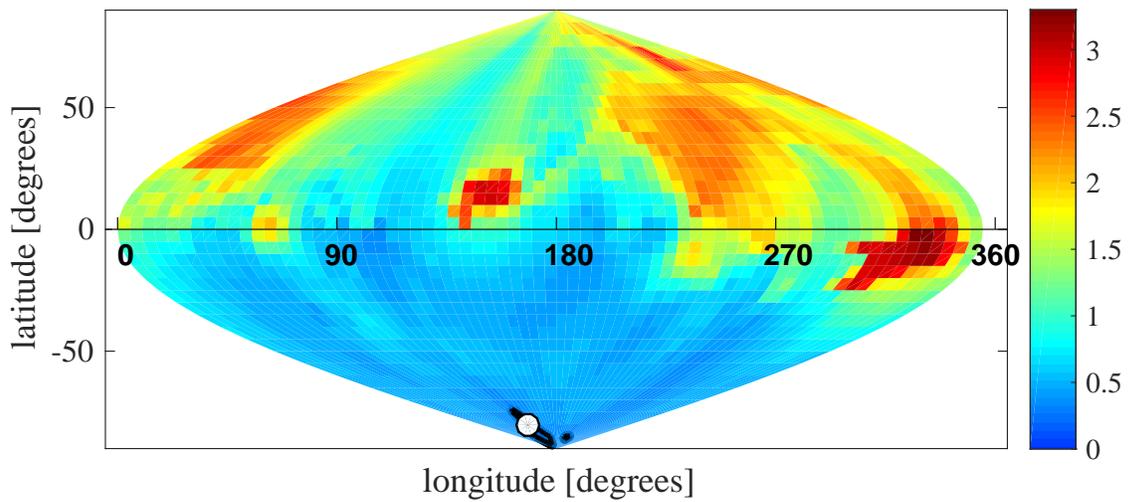}}
\caption{The $\chi^2$ values for all spin axis solutions on a longitude-latitude plane for {\it OE84}. The uncertainty of the fit corresponding to $2\sigma$ (black line) above the global minimum clearly demonstrates the retrograde sense of rotation. The global best-fit solution has $\chi^2 = 0.27$ (white circle) and it indicates a spin axis pointed close to the ecliptic south pole.
\label{fig:OE84_Chi2Pole}}
\end{figure}

\begin{figure}
\centerline{\includegraphics[width=17cm]{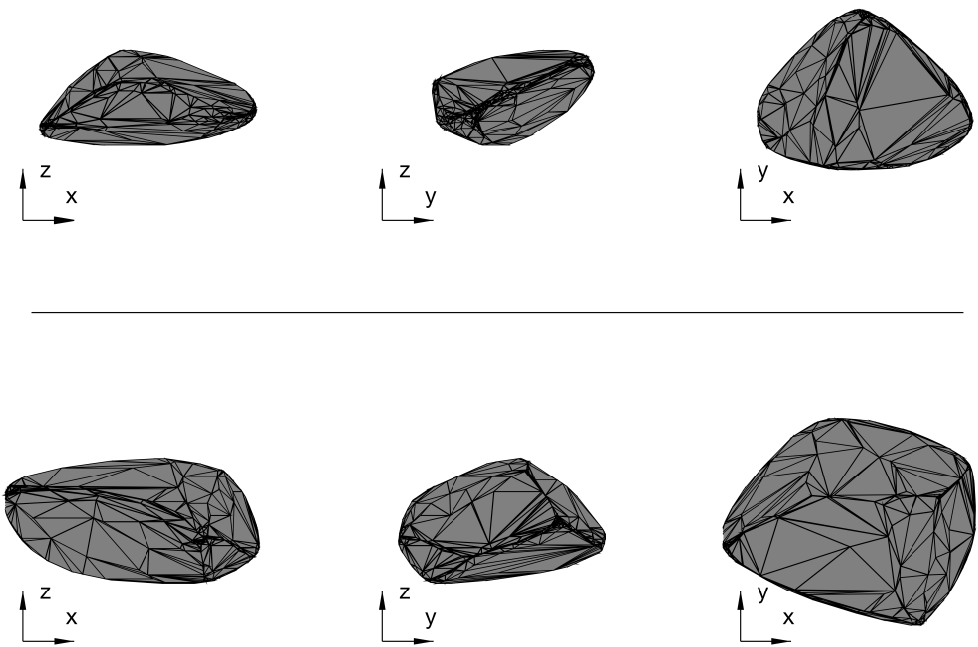}}
\caption{Two possible shape models for {\it OE84}, one shape per row. The shapes are plotted from two perpendicular side views (left and middle panels) and from a top view (right panels).
\label{fig:OE84_BestShape}}
\end{figure}

We found that the combined photometric data of {\it OE84} from the 2001 and 2016 apparitions are best matched to 14 solutions that represent four sidereal periods, two shape models, two spin axis longitude and one spin axis latitude. Each of these models were statistically significant at the $2\sigma$ level ($>95\%$, Fig.~\ref{fig:OE84_Chi2Pole}). The differences across values for same parameter are minor, hence, we give the mean of each parameter, with the standard deviation as the uncertainty: a sidereal period of $0.486542\pm0.000002$~h, a spin axis longitude of $115\pm5^o$ or a $229\pm5^o$, retrograde sense of rotation with a spin axis that is close to the ecliptic south pole (ecliptic latitude of $-80\pm5$ degrees) and physical axis ratios of $A/B=1.32\pm0.04$ and $B/C=1.8\pm0.2$. The two shape models are presented in Fig.~\ref{fig:OE84_BestShape}. Adding a linear parameter for a spin-rate change, did not improve the $\chi^2$ of the best solutions, therefore we conclude that the asteroid spin rate remained fixed in the last 14 years within the error range of $2\cdot10^{-6}~$hours. We note that a pole position very close to the ecliptic normal argues in favor of past YORP evolution for {\it OE84}, since YORP spin-up adds angular momentum normal to the ecliptic (Vokrouhlick{\'y} et al. 2003, Hanu{\v s} et al. 2011).

An alternative way to approximate the adjustment of synodic to sidereal periods within an individual apparition is based on the change in the angle from which the asteroid is observed from the Earth. Harris et al. (1984) defined the Phase Angle Bisector (PAB), which is the mean of the geocentric and heliocentric vectors to the asteroid, and its motion can indicate the difference between the synodic and the sidereal period. We applied this technique separately on the 2001 and 2016 data sets while assuming a retrograde sense of rotation as indicated from the lightcurve inversion technique. During 2016, the changing geometry of the asteroid, Earth, and Sun resulted in a small positive difference in PAB longitude (Table~\ref{tab:ObsCircum}). This produced a small synodic to sidereal adjustment of $0.00025~$seconds. During the 2001 observations, a larger positive difference in PAB longitude resulted in a synodic to sidereal adjustment of $0.056~$seconds.

Using these adjustments we calculate the sidereal periods from 2001 and 2016, and find a difference of $0.02$ to $0.09~$seconds between them (this includes the uncertainty on the synodic periods). However, this is the same order of magnitude as the adjustment from synodic to sidereal period for the 2001 data ($0.056~$seconds). Therefore, we dismiss it as an insignificant difference attributable to the sensitivity of this method to the roughness of the surface and detailed changes in viewing geometry. Asteroid surfaces are not perfect light scatterers or geometrically symmetric, thus their unknown roughness parameter can add large uncertainty to the synodic-sidereal adjustment using the PAB procedure (Pravec et al. 2005).

Below we investigate the implications of a zero spin difference between 2001 and 2016, and how this constrains the physical nature of this unique asteroid.

\section{Discussion}
\label{sec:discussion}

Our 2016 photometric observations of {\it OE84} confirm its unique fast rotation as measured by Pravec et al. (2002) almost a decade and a half later. In about half an hour this $\sim650~$m body completes one rotation while other similar-sized asteroids spin at least 4 times slower. Here we test analytically what is the cohesion needed to keep a rubble pile {\it OE84} from falling apart and shed mass.

\subsection{A rubble-pile with high cohesion}
\label{sec:StrongCohesion}

The Drucker-Prager yield criterion calculates the shear stress in a rotating ellipsoidal body at breakup, hence, it allows us to constrain the body's cohesion. This criterion requires a simplified approach that treats {\it OE84} as a homogenous object with fixed density throughout its interior. We applied this criterion, adapted from Holsapple (2004, 2007) and Rozitis et al. (2014), to the measured parameters of {\it OE84} (summarized in Table~\ref{tab:PhysicalParam}) and assuming a bulk density\footnote{This is the density of asteroid (25143) Itokawa (Fujiwara et al. 2006, S{\'a}nchez and Scheeres 2014), which is similar in size to {\it OE84}, has the same spectral classification, and has a known rubble pile structure.} of $2000~$$\rm{kg~m^{-3}}$ and an angle of friction\footnote{The mean friction angle value of geological materials (Hirabayashi \& Scheeres 2015).} of $35^{o}$. Using the mathematical formalism described in Polishook et al. (2016) we derived a minimum possible cohesion of $1080^{+790}_{-420}~$Pa.

This value is significantly higher than cohesion values estimated for rubble pile asteroids. These range from $\sim25~$Pa based on the grain size distributions of (25143) Itokawa and active asteroid {\it P/2013 P5} (S{\'a}nchez and Scheeres 2014) to $64^{+12}_{-20}~$Pa constrained by a detailed physical model (that includes radar observed size and shape, thermal inertia, and bulk density) of the fast rotator (29075) {\it 1950 DA} (Rozitis et al. 2014; though Gundlach \& Blum (2015) suggest the value is between $24$ and $88~$Pa). Other constraints, that are based on asteroids with fewer measured parameters and therefore give cohesion values with high uncertainties, include $40-210~$Pa for the precursor body of the active asteroid {\it P/2013 R3} (Hirabayashi et al. 2014), $150-450~$Pa constrained by the spin rate and shape of (60716) {\it 2000 GD65} (Polishook et al. 2016) and $\sim100~$Pa constrained solely by the spin rate of (335433) {\it 2005 UW163} (Polishook et al. 2016). 

If indeed the cohesion of {\it OE84} is as high as this model suggests, then this asteroid would be unique with an estimated cohesion up to an order of magnitude greater than other comparable objects. This is most probably not due to compositional differences, since {\it OE84} has an S-type taxonomy like many other similar-sized NEAs (Binzel et al. 2014). Another possibility is that the high cohesion is due to unusually small grains that push the cohesion to high values. Indeed, the calculated cohesion value for {\it OE84} is similar to the top end of the cohesion range of lunar regolith, sampled by Apollo astronauts (Mitchell et al. 1974), that extends from $100$ to $1,000~$Pa. Small-sized grains might exist on {\it OE84} if it is an older body that, compared to other asteroids, suffered from more non-catastrophic collisions, micrometeorite bombardment, Brazil nut effect (Perera et al. 2016), and thermal fracturing (Delbo et al. 2014). Such a history may have fractured {\it OE84}'s internal structure, reduced the size of grains within its interior and on its surface, and thereby increased its cohesion. However, the collisional lifetime of asteroids in the size range of {\it OE84} is in the order of $10^8~$years (Bottke et al. 2005), much younger than the age of lunar regolith. Therefore, this high-cohesion scenario demands an old {\it OE84} compared to an average asteroid in this size range. This raises questions about the viability of this model.

Another argument against this model is based on the assumption that van der Waals forces are the source of asteroids' cohesion. In this case, we can estimate the mean radius $\bar{r}$ of the grains that cause boulders to stick together (S{\'a}nchez \& Scheers 2014):

\begin{equation}
\sigma = 1.56 \cdot 10^{-4} / \bar{r} ~~~[Pa],
\label{eq:grainSize}
\end{equation}

where $\sigma$ is the tensile strength of a randomly packed matrix of grains, which is equal to the cohesion in the case where no pressure is present (S{\'a}nchez \& Scheers 2014). This gives an average grain size of $\sim10^{-7}~$m which is two orders of magnitude smaller than the estimated mean grain size of (25143) Itokawa and active asteroid {\it P/2013 P5} ({\it e.g.}, S{\'a}nchez \& Scheers 2014). This is also at least an order of magnitude smaller than the smallest particles in the Hayabusa sample chamber returned from Itokawa (Nakamura et al. 2011).

An alternative calculation for the minimal cohesion that requires smaller cohesion was presented by Hirabayashi et al. (2015). They suggested a rubble pile model with a core that is still composed of discrete components but is significantly stronger than the shell around it. This is based on numerical simulations which showed that a rubble pile with a uniform distribution of material properties will be prone to internal fracture before mass can be shed from the surface ({\it e.g.}, Hirabayashi 2015, Hirabayashi \& Scheeres 2015, S{\'a}nchez 2016). This led Hirabayashi \& Scheeres (2015), for example, to constrain a cohesive strength of at least $75-85~$Pa for the internal core of the fast rotator (29075) {\it 1950DA}.

\begin{figure}
\centerline{\includegraphics[width=17cm]{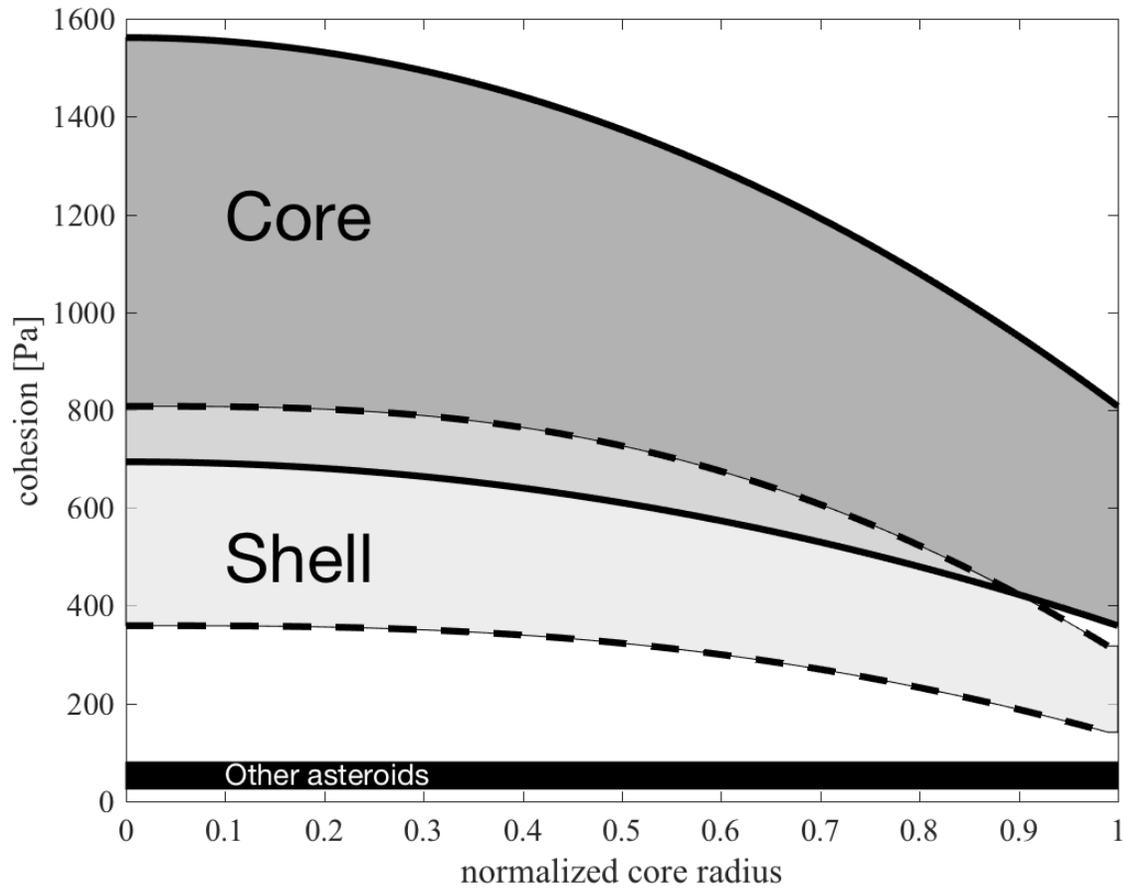}}
\caption{The expected lower limit on the cohesion of the core (dark grey range within solid lines) and the shell (light grey range within dash lines) as a function of the core's radius, normalized by the asteroid radius. The core's cohesion at $100\%$ radius is equal to the shell's cohesion at $0\%$ radius. The possible minimal core cohesion within the uncertainty range is $\sim400~$Pa, when the core is almost as large as the asteroid, which is 5 to 16 times larger than the cohesion estimation for other asteroids ($25$ to $80~$Pa; black rectangle at the bottom).
\label{fig:OE84_CoreCohesion}}
\end{figure}

We follow the analytical model presented by Hirabayashi et al. (2015; Eq. 8 to 22) and apply it to {\it OE84}. We derive a wider range of values for the minimal cohesion: between $\sim400$ to $\sim1600~$Pa (Fig.~\ref{fig:OE84_CoreCohesion}). While the lower edge of this range is smaller than the homogenous rubble pile described above, it is still $\sim5$ to $16$ times higher than the cohesion estimates for other asteroids. Moreover, the Hirabayashi et al. model assumes for simplicity a spherical body. Since {\it OE84} is elongated it probably requires a higher cohesion value than the analytic constraint.

A highly cohesive core is possible if we assume that grains within asteroid interiors are smaller, and therefore more cohesive, than grains on the surface. This might be caused by the so-called Brazil nut effect, which segregates particles of different sizes with floatation of larger fragments above smaller ones (Tancredi et al. 2015). However, Perera et al. (2016) have found that while size segregation can take place near the surface, the innermost regions remain unsorted under even the most vigorous shaking. Alternatively, a stronger core is possible if the constituent grains that comprise it are significantly more compact, therefore producing higher bulk densities and a stronger core. However, {\it OE84} is only $\sim650~$m across, and since this model assumes a rubble pile structure throughout (albeit a non-uniform structure), it is not clear how the interior of a body of this size, with such low self gravity, would compact. Asteroid structures compact in bodies larger than $100~$km across (Richardson et al. 2002), about 3 orders of magnitude larger than {\it OE84}. Thus, we suggest that the asteroid is most plausibly not a rubble pile with significant cohesion.

\subsection{A monolithic body}
\label{sec:monolith}

Alternatively, a monolithic structure for {\it OE84} would resist disruption since the cohesion of solid rocks is on the order of $10^6~$Pa (S{\'a}nchez 2016). In this case {\it OE84} could be a ``bald'', boulder-less body. While we cannot exclude such a scenario, it seems unlikely since all {\it in-situ} and radar observed asteroids, larger than $\sim200~$m, including similar-sized (25143) Itokawa, have boulders and regolith on their surfaces ({\it e.g.} Benner et al. 2015). Moreover, if {\it OE84} was formed in a catastrophic collision from a much larger parent body, then it is expected that debris from that collision would reside on its surface (Mazrouei et al. 2014). Therefore, we test a more likely model of a monolithic body with gravel and boulders scattered on its surface.

Boulders laying on top of a monolithic body rotating as fast as {\it OE84} would not remain on the surface through self gravity alone -- a cohesive force is needed. Simplifying the forces acting on a boulder with radius $r$ and a mass $m$ (Hirabayashi \& Scheeres 2015), the boulder will remain on the surface if

\begin{equation}
m(R+r)\omega^2 \leqslant \frac{GMm}{(R+r)^2} + f_{\rm{cohesion}},
\label{eq:boulderCohesion}
\end{equation}
where $\omega$ is the asteroid's spin rate, $M$ is its mass, and $R$ is its radius. $ f_{\rm{cohesion}}$ is the cohesion force which can be simply defined as $\sim2k\pi r^2$, where $k$ represents cohesion and $\sim2\pi r^2$ is the apparent area of the boulder touching the surface. Using relevant values for {\it OE84}, a boulder density of $3300~\rm{kg~m^{-3}}$ (averaged density of ordinary chondrite meteorites, the analogs for the S-type asteroids; Carry 2012) and a cohesion value of $25~$Pa (S{\'a}nchez and Scheeres 2014), the diameter of surface boulders would be limited to $\leqslant5-7~$m. Larger boulders would need larger cohesion. This is almost one order of magnitude smaller than the largest boulder on (25143) Itokawa ($\sim40~$m; Michikami et al. 2008), and (4179) Toutatis ($>50~$m; Jiang et al. 2015). A cohesion value as large as $85~$Pa, which is the maximal cohesion range estimated for (29075) {\it 1950 DA} (Gundlach \& Blum 2015, Hirabayashi \& Scheeres 2015), allows boulders of $15$ to $23~$m on the surface of {\it OE84}, which are the same order of magnitude as the largest boulders on (25143) Itokawa and (4179) Toutatis. These values suggest that a number of boulders with a few meters to a few tens of meters in size, would be possible for {\it OE84}. This would be consistent with the physical structures of asteroids with which we are currently familiar.

\subsection{Quantifying the uniqueness and age of {\it OE84}}
\label{sec:uniqueness}

The location of {\it OE84} on the size-spin diagram (Fig.~\ref{fig:DiamSpinDiag}) and our analytical calculations (previously in this section) suggest that {\it OE84} is unique in one (or more) of its internal physical properties and/or experienced a unique history. Before we decide whether these may be the case, we estimate how unique {\it OE84} is relative to other NEAs. Currently, there are only two other objects in the size range of {\it OE84} (144977 and 335433; Chang et al. 2017, 2014, respectively) confirmed to rotate significantly faster than the $\sim2~$hours spin barrier. Four other fast-rotating candidates need to be confirmed (Chang et al. 2015), setting the number of these unique asteroids to at least 3 to 7. Since an unusually high cohesion or a monolithic structure does not necessarily imply fast rotation, we can assume that in this size range there are additional asteroids with slow rotation that have the same unusual physical structures. This allows us to constrain the minimal number of {\it OE84}-like structured asteroids to $3-7$ out of the 1390 asteroids\footnote{1390 is the number of asteroids with diameters in a factor of 2 of {\it OE84}'s diameter and a known rotation period. Data is from Warner et al. (2009).} in this size range. Hence, a few tenths of a percent of the current catalog, are similar to {\it OE84} in terms of internal structure or history.

If indeed $\it OE84$ has a monolithic structure, we need to explain why most other asteroids do not. The canonical view is that monolithic bodies are formed in parent bodies large enough ($>100~$km; Richardson et al. 2002) to compact material and lithify their interiors. Following a destructive collision, monolithic fragments are free to orbit the sun ({\it e.g.} Elkins-Tanton et al. 2011) but with time will suffer multiple destructive collisions and re-accretion events until they become rubble piles. Therefore, a monolithic structure for {\it OE84} indicates a young age compared to the other $99.9\%$ asteroids that are of a similar size. Using the intrinsic collision probabilities between NEAs from Bottke et al. (1994), we can estimate the number of impacts a given NEA or a main belt asteroid suffer each year:

\begin{equation}
N_{\rm{impacts}} = P_i N (> r_{\rm{projectile}}) (r_{\rm{OE84}} + r_{\rm{projectile}}) ^ 2,
\label{eq:collisionProbability}
\end{equation}

where $r_{\rm{OE84}}=325^{+80}_{-55}~$m and $P_i = 15.3e-18~\rm{km^{-2}~yr^{-1}}$ for NEA and $P_i = 2.85e-18~\rm{km^{-2}~yr^{-1}}$ for a main belt asteroid. We chose $r_{\rm{projectile}}=10~$m since it was shown that similar size impactors can crack kilometer sized  bodies (Asphaug et al., 1998; Asphaug and Scheeres, 1999). We calculate that an {\it OE84}-sized asteroid will suffer such a collision, on average, every few $10^7~$years (for NEA) or every $10^8~$years (for main belt asteroid). This indicates that {\it OE84} would have to be younger than a few times $10^7~$years.

The age of {\it OE84} can also be constrained from its source region as a NEA. Based on its orbital elements and using source regions probabilities from Bottke et al. (2002), there is $74\%$ that {\it OE84} originated from the Mars crossers population and $18\%$ that it passed through the {\it $\nu 6$} resonance. This agrees with its retrograde sense of rotation, since the asteroid will have to move via the Yarkvosky effect from the main belt inward, reducing its semi-major axis, in order to enter these resonances. The mean dynamical lifetime of NEAs from these source regions is $3.75$ and $6.54~$Myrs, respectively. Since this timescale is shorter than the collisional timescale calculated above, it means that most NEAs do not suffer a collision while being NEAs, thus {\it OE84} is not unique in this manner. Therefore, in order to keep its unique monolithic structure for this model, {\it OE84} should have been removed from the main belt very soon ($\lessapprox10^7~$years) after it was liberated from its parent body.

Since the pole position argues in favor of YORP evolution, an additional way to constrain the age of {\it OE84} is by considering the timescales associated with YORP spin-up of the body ({\it e.g.} Rozitis \& Green 2013, Jacobson et al. 2014). If the initial rotation period of {\it OE84} was on the $2.2~h$ (the spin barrier), then YORP can spin-up the asteroid to $P\sim0.5~h$ in $2$ to $50$ million years. This range sets a minimal constraint on {\it OE84}'s age and also suggests that {\it OE84} has been removed from the main belt within $\lessapprox10^7~$years, as stated in the previous paragraph.

\section{Summary and conclusions}
\label{sec:conclusions}

We observed the near-Earth asteroid {\it (455213) 2001 OE84} 14 years after its fast spin was first measured by Pravec et al. (2002). We confirmed {\it OE84}'s fast rotation of $\sim0.5~$hour and could not identify a period alteration within the uncertainty of the sidereal period. Using the lightcurve inversion technique, we found that {\it OE84} has a retrograde sense of rotation with the spin axis very close to the ecliptic south pole.

We showed that {\it OE84} is rotating too fast to remain bound as a uniform rubble pile with cohesion values derived for other asteroids. A non-uniform rubble pile model, where the central region is stronger than the shell, would also break-up with cohesion values derived for other asteroids. Only a monolithic body can rotate as fast as {\it OE84}. We also demonstrated, using a simplistic model, that relevant cohesion values can retain boulders on the surface as large as those observed on other asteroids visited by spacecraft. Constraints applied from intrinsic collision probabilities and the evolution of spin due to the YORP effect, set an age for {\it OE84} between $2-10$ million years, as the time since it was liberated from its parent body.

The existence of a shell of boulders, the possibility of a bald monolithic body, or a high cohesion value for {\it OE84}, may be determined in the future with thermal infrared observations to measure the thermal inertia of the surface. Bare rock has thermal inertia larger than $2500~\rm{J~m^{-2}~s^{-0.5}~K^{-1}}$, while the regolith covered lunar surface has a value of $50~\rm{J~m^{-2}~s^{-0.5}~K^{-1}}$ (Delbo et al. 2007). Similar-sized (25143) Itokawa, that is covered with thousands of boulders and gravel, has a thermal inertia of $\sim700~\rm{J~m^{-2}~s^{-0.5}~K^{-1}}$ (Muller et al. 2014). Infrared observations could also provide a measured albedo and not a value based on the mean albedo of {\it OE84}'s spectral classification. In addition, near-IR spectroscopy could inform the specific mineralogy of the asteroid and thus enable ties to specific collisional families in the main belt, especially those located near the $\nu_6$ resonance. A possible suspect could be an as yet unidentified young sub-family within the Flora cluster. This could provide additional constraints on the age of the object. The next favorable observing conditions for {\it OE84} are August 2025 and October 2032.

While {\it OE84} has a unique combination of size and spin rate that may be the case for only one in a thousand asteroids of {\it OE84}'s size, mitigation plans against asteroid impacts would benefit by considering the possibility of solid monolithic bodies within the NEA population.

\acknowledgments

We thank Masatoshi Hirabayashi for fruitful discussions about his asteroid structure model, Josef Durech for the code of the lightcurve inversion technique and Petr Pravec for providing us the measurements from 2001. We thank the referees, Alan Harris (Cal-Al) and Josef Hanus, for their constructive comments. DP is grateful to the ministry of Science, Technology and Space of the Israeli government for their Ramon fellowship for post-docs. NM and AT acknowledge support from NASA NEOO grant number NNX14AN82G awarded to the Mission Accessible Near-Earth Object Survey (MANOS). OA acknowledges support from the Helen Kimmel Center for Planetary Science, the Minerva Center for Life Under Extreme Planetary Conditions and by the I-CORE Program of the PBC and ISF (Center No. 1829/12). Sansa Stark is thankful to Lord Petyr Baelish for providing the Vale army. These results made use of Lowell Observatory's Discovery Channel Telescope. Lowell operates the DCT in partnership with Boston University, Northern Arizona University, the University of Maryland, the University of Toledo and Yale University. Partial support of the DCT was provided by Discovery Communications. LMI was built by Lowell Observatory using funds from the National Science Foundation (AST-1005313). We are grateful to the people of Arizona that support hosting professional observatories and for endorsing astronomical studies.

\begin{deluxetable*}{ccccccccc}
\tablecolumns{9}
\tablewidth{0pt}
\tablecaption{Circumstances of observations}
\tablehead{
\colhead{Date} &
\colhead{Filter} &
\colhead{Time span} &
\colhead{Image Num.} &
\colhead{${\it r}$} &
\colhead{${\it \Delta}$} &
\colhead{${\it \alpha}$} &
\colhead{$L_{PAB}$} &
\colhead{$\beta_{PAB}$} \\
\colhead{}          &
\colhead{}          &
\colhead{[hour]}    &
\colhead{}          &
\colhead{[AU]} &
\colhead{[AU]} &
\colhead{[deg]} &
\colhead{[deg]} &
\colhead{[deg]}
}
\startdata
19 Jan 2016 & VR & 3.06 & 43 & 2.00 & 1.16 & 18.97 & 145.7 & 11.5 \\
16 Feb 2016& VR & 2.03 & 65 & 2.16 & 1.20 & 7.03   & 148.0 & 11.0 \\
10 Mar 2016 & VR & 0.79 & 32 & 2.29 & 1.38 & 12.75 & 149.5 & 10.1 \\
12 Mar 2016 & VR & 1.25 & 58 & 2.30 & 1.40 & 13.47 & 149.6 & 10.0
 \enddata
\tablenotetext{}{Columns: date of observation, filter, hourly time span, number of images, heliocentric and geocentric distances, phase angle, and Phase Angle Bisector (PAB) ecliptic longitude and latitude.}
\tablenotetext{}{All observations were conducted using Lowell Observatory's 4.3m Discovery Channel Telescope.}
\label{tab:ObsCircum}
\end{deluxetable*}

\begin{deluxetable*}{lll}
\tablecolumns{3}
\tablewidth{0pt}
\tablecaption{physical parameters of (455213) {\it 2001 OE84}}
\tablehead{
\colhead{Parameters} &
\colhead{Values} &
\colhead{Reference}
}
\startdata
Absolute Magnitude $H_V$ & $18.31\pm0.16~$mag & Pravec et al. (2002). \\
Spectral classification& S-complex & Pravec et al. (2002). \\
Albedo $P_v$ & $0.197\pm0.051$ & Average value for S-complex asteroids (Pravec et al. 2012). \\
Diameter D & $650^{+160}_{-110}~$m & $D=\frac{1329}{\sqrt{P_V}}10^{-0.2H_V}$ \\
Synodic Rotation Period & $0.486545\pm0.000004~$hours & Our photometric observations \\
Sidereal Rotation Period & $0.486542\pm0.000002~$hours & The lightcurve inversion technique \\
Mean lightcurve amplitude & $0.45\pm0.05~$mag and $0.28\pm0.03~$mag & Our photometric observations \\
Triaxial ratios & $A/B=\sim1.32\pm0.04$ & The lightcurve inversion technique \\
Triaxial ratios & $B/C=\sim1.8\pm0.2$ & The lightcurve inversion technique \\
Spin axis longitude & $\lambda=115^o$ or $229^o$ & The lightcurve inversion technique \\
Spin axis latitude & $\beta=-80^o\pm5^o$ & The lightcurve inversion technique
\enddata
\label{tab:PhysicalParam}
\end{deluxetable*}

\end{document}